# EXPERIMENTAL AND NUMERICAL STUDIES OF INTERACTION BETWEEN LIQUID DROPLETS AND HOT SURFACES UNDER DIFFERENT CONDITIONS

Subhasish Guchhait[#1], Pratim Kumar[2]

[1] PhD Scholar, Department of Mechanical and Aerospace Engineering, IIT Hyderabad

[2] Assistant Professor, Department of Aerospace Engineering and Applied Mechanics, IIEST Shibpur [E-mail: Pratim.kumar.86@gmail.com]

## ABSTRACT

In the present paper, experimental and numerical studies of interaction between different liquid droplets with different hot metal surfaces had been carried out and the obtained results were interpreted using graphs and pictures. Droplet impact, spreading, break-up, and leidenfrost phenomenon were observed with the help of video camera at 100 fps. The interaction studies were done at different surface temperatures and different droplet impinging velocities. Three metals and three liquids were selected for the present study. Copper, Aluminium and Steel were selected as metals with different roughness factors and at different temperatures, while in liquid entities, Methanol, Ethanol and Kerosene (Water was also used in some portion) were selected. The selections of the metals were based on thethermal diffusivity property as it influences the thermal interaction between droplet and the hot metal surface. The temperature range of metal surface was kept from 150° C to 425°C and the impinging velocity range of droplets were in the range of 2.4 - 4.9 m/s. For computational studies Ansys Fluent R3-2019 version was used. Numerical study for the interaction process was carried out using Volume of Fluid (VOF) method, solving Navier-Stokes and energy equations in a 2D-geometry transient simulation. Parameters like surface temperature, heat flux, and spreading ratio with respect to Weber number change were investigated and discussed in the present paper. The output of this study will be useful to understand the physical mechanisms better and to develop novel models for effective spray cooling system designs. The results show that for a lower value of Weber number, after impact the droplets spread less wide compare to high Weber number. The spreading ratio of droplets was almost constant or uniform for different surface temperature and it depends only on the surface roughness.

1. **INTRODUCTION**

Two-phase flows carrying droplets are very common in nature and engineering. Nowadays many researchers investigate the interaction between droplets and solid walls in details as these processes have considerable influence on two-phase flows [1-9]. Some common examples are fuel droplets in combustion processes, ink jet printing and inhalation of medical sprays. Spray technology plays an important role in cooling down the heated surfaces in many industrial applications such as emergency cooling systems of aerospace engine components, supercomputers, reactors of nuclear power plants etc [10-11]. The mechanism of extracting heat from the hot surfaces using spray cooling method are highly complicated as this technique depend on many parameters like spray, surface, and fluid characteristics [12-14]. These highly complicated heat and mass transport mechanisms between droplet and the heated surface interaction is of very short duration of time i.e. in microseconds. Surface temperature plays an important role in this interaction study as it used to influence heat transfer performance and impact dynamics. Therefore, it is very important to understand the phenomenon of droplet impact on the heated walls.

According to the experimental investigation there can be six possible regimes for droplet after impacting on solid surfaces and they are: deposition, splashing, receding breakup and rebound. Each of these regimes is dependent upon impact velocity, liquid surface tension, droplet diameter, surface temperature, viscosity etc [15-16]. **Deposition** states that the droplet stays attached to the surface after deforming. At the beginning of spreading of droplet, tiny droplets are generated at the contact line; it is known as **Prompt splash**. **Corona splash** occurs when droplets are formed around the rim of a corona, remote from the solid surface. When the tiny droplets are left behind by the receding lamella during retraction process that phenomenon is known as **Receding breakup**. **Partial rebound** and **complete rebound** happen when the drop spreads with a relatively large spreading diameter and recedes with a relatively high receding contact angle. In the complete rebound, the receding contact angle is even lager to form a very energetic impact.

Surface temperature is an important characteristic affecting the dynamics of impacting droplets, particularly when heat transfer is a concern. In the case of a hot, dry solid surface, the droplet impact outcomes are differentiated in different regimes evaporation, nucleate boiling, foaming, transitional boiling, and film boiling. **The nucleate boiling regime** occurs at relatively high surface temperatures. In this process the droplet is in direct contact with the surface. Also the vapour bubbles are formed at various isolated nucleation sites. **The**

**foaming** is a subcategory of nucleate boiling. In this regime the entire drop starts to foam & the vapour bubbles grow much larger while no separation from the liquid–gas interface. **In the transition boiling regime**, the generation rate of the vapor bubble increases quickly due to high wall temperature. In this regime, liquid layers frequently collapse, as a result it is very unstable and secondary droplets are also generated. If the temperature is further increased then the regime will convert to **film boiling**. In this regime a vapor layer forms in between the droplet and the solid surfaces & as a result the droplet floats on the vapor layer, here the vapour layer acts as a thermally insulating film and due to it the evaporation rate becomes more slowly than if it remained in contact with the surface. This phenomenon is known as the **Leidenfrost effect**; and the temperature at which the droplet evaporation time reaches its maximum is called the **Leidenfrost temperature**.

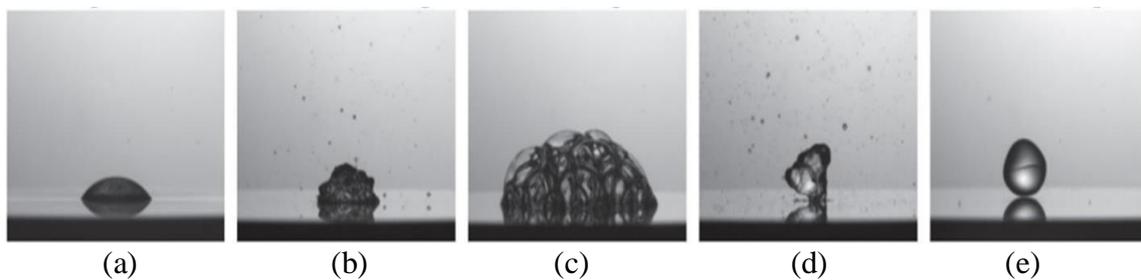

(a)　　　　(b)　　　　(c)　　　　(d)　　　　(e)

*Figure-1: Outcomes of droplet collision with a hot, dry solid surface: (a) evaporation, (b) nucleate boiling, (c) foaming, (d) transition and (e) film boiling* [17]

In 1876 the phenomenon of droplet impact experiment was studied by Worthington from then the topic continues to influence the research scholars of various disciplines [18]. This experiment gave the emphasis on droplet spreading, splashing, receding, bouncing, foaming, bubble entrapment and atomization after droplet impact on hot surfaces [19]. Surface tension, inertial force, roughness, viscous force and wettability of solid surfaces affect the droplet after impact while the heat transfer efficiency between the wall and the droplet is directly affected by the impact dynamics. According to Alizadeh et al. [20] wall temperature can influence the spreading and retraction behaviours of droplets. Jin et al. [21-22] studied the droplet behaviour impacts on cold surface. The result was totally different with respect to the result of impact on hot temperature surface. After spreading to its maximum spreading diameter the droplet would not retract and it would freeze faster on super cooled surface with increasing in Weber number. Many researchers have investigated the droplet spreading by numerical simulation e.g., Lattice Boltzmann method [23-26], level set method [27] etc. Using a modified volume-of-fluid (VOF) method Bussmann et al. simulated the splashing and fingering of a droplet after impact on a solid surface [28]. Villegas et al [29] used a ghost-fluid level-set method to simulate the impact of a drop where the surface temperature was above the Leidenfrost point. In their work, they had only studied about the rebound of a drop for Weber number less than 60. Bernadin et al. [30] stated that the impact Weber number and surface temperature are the main factors governing impact behavior and heat transfer. Erkan et al. [31] developed experimentally validated a new numerical method (VOF-MPS) based on the moving particle semi-implicit (MPS) to investigate the droplet deposition onto liquid film. They compared this method with MPS-CSF method. They stated that VOF–MPS method predicts temporal variation of the crown diameter with a small deviation from the experiments.

In the present paper, we had investigated experimentally and numerically the important characteristics which affect the cooling performance for different surface temperature and Weber number for different liquid. The ultimate goal was to show the droplet shape after impact on the hot surface and analysis the factors that affect the evaporation process. Finally the experimental results were compared with the numerically obtained ones and discussed elaborately.

## 2. EXPERIMENTAL AND NUMERICAL SIMULATIONS METHODOLOGY

In this chapter, two methodologies which were used for studying the droplet dynamics with hot surfaces have been discussed in elaborately. The two approaches are Experimental and Numerical methodology. Each methodology was discussed separately in section 2.1 and in section 2.2 respectively.

### 2.1 EXPERIMENTAL METHODOLOGY

A schematic view of the experimental set-up is shown in Figure-2. We had used three different surfaces and those are Copper, Aluminium & Steel and for droplet generation Methanol, Ethanol and Kerosene were used. By using of the roughness tester [Haridarshan Instru-Lab, India] firstly the surface roughness value of each metal plate had been tested as it played a vital role for the spreading diameter of the droplet [9]. The set-up stand, camera and the flash light were placed according to Figure-2. The impinging velocity of the droplet could be changed by changing the injector height. For high impinging speed volume ratio of droplets became very less and also the loss of momentum became high which affects the evaporation rate of the droplet. After pouring the liquid into the injector, the heater was powered on so that the temperature of the surface could increase. The thermocouple (K-Type) was kept in such a manner so that it could give accurate temperature reading.

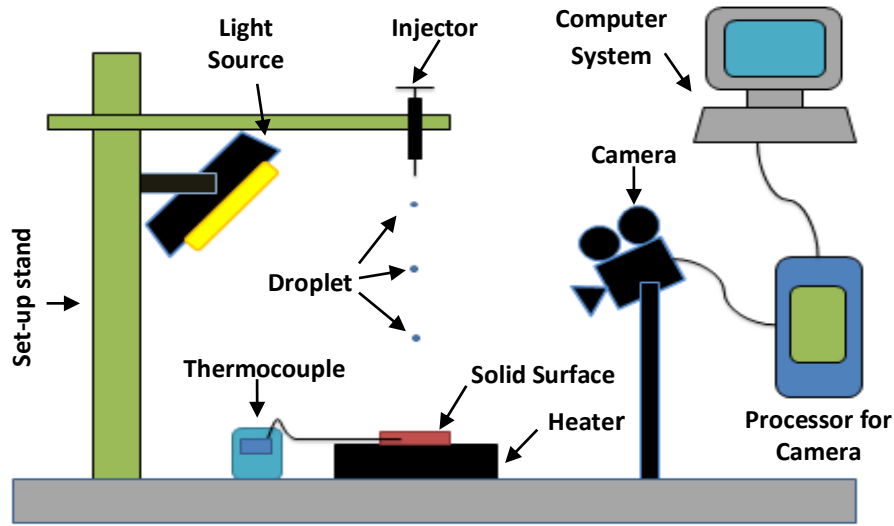

*Figure-2: Sketch of complete experimental set-up*

After starting the recording, the injector was pushed in a very gentle way to release a single droplet. The video recording was kept switched on till the evaporation of the droplet completed. Same procedure was done for each liquid on each surface at different temperature and different impinging velocity (2.4, 3.9 & 4.9m/s). To know the effects of surface temperatures, we had used three different temperatures (150°C, 300°C & 425°C) for every surface. For high temperature, volume ratio of droplets becomes low and the evaporation effect becomes high. The computer system was used to extract the pictures from the video so that we got the images of different droplet characteristic after impact by which modeling of different physical phenomena were done which may include: wetting area or deposition, splash, receding, rebound, leidenfrost effect, levitation of droplet, boiling etc.

## 2.2 NUMERICAL METHODOLOGY

To track the interface of droplet surface interaction the volume of fluid method (VOF) was used. For both liquid and vapour source terms had been included in the momentum, mass and heat transfer. Transient simulation was used as the condition wasn't steady.

$$\nabla . (\rho u) = \dot{\rho} \quad \text{-----------------------------------------------} \quad (1)$$

$$\frac{\partial \rho u}{\partial t} + \nabla . (u . \rho u) = -\nabla p + \nabla . (\mu . \nabla u) + f_\sigma + g \quad \text{----------------------} \quad (2)$$

$$\frac{\partial \rho c T}{\partial t} + \nabla . (u . \rho c T) = \nabla . (k . \nabla T) + \dot{h} \quad \text{-------------------------} \quad (3)$$

In this model dissipation term in the energy equation was ignored and liquid and gases were assumed to be incompressible. To capture the interface an additional equation for volume fraction was solved.

$$\frac{\partial \alpha}{\partial t} + \nabla \cdot (u.\alpha) = \frac{\dot{\rho}}{\rho}\alpha \quad \text{------------------------------------} \quad (4)$$

Here α is volume fraction and it is defined by the ratio of Volume of liquid phase to total control volume. The value of α is 1 inside the liquid and 0 in the gas phase and lies between 0 and 1 in the interface area.

$$\rho = \alpha \rho_{liq} + (1-\alpha)\rho_{gas} \quad \text{------------------------------} \quad (5)$$

$$\mu = \alpha \mu_{liq} + (1-\alpha)\mu_{gas} \quad \text{------------------------------} \quad (6)$$

The above equations were solved in the commercial software package of ANSYS Fluent R3-2019. A volume fraction diagram of a single droplet with the boundary conditions of the domain was illustrated in Figure-3. Pressure outlet conditions were applied on top, right and left sides of the domain and in the bottom domain constant temperature boundary condition was applied. To solve the simulation explicit method was used and k-ω-SST turbulence model is applied.

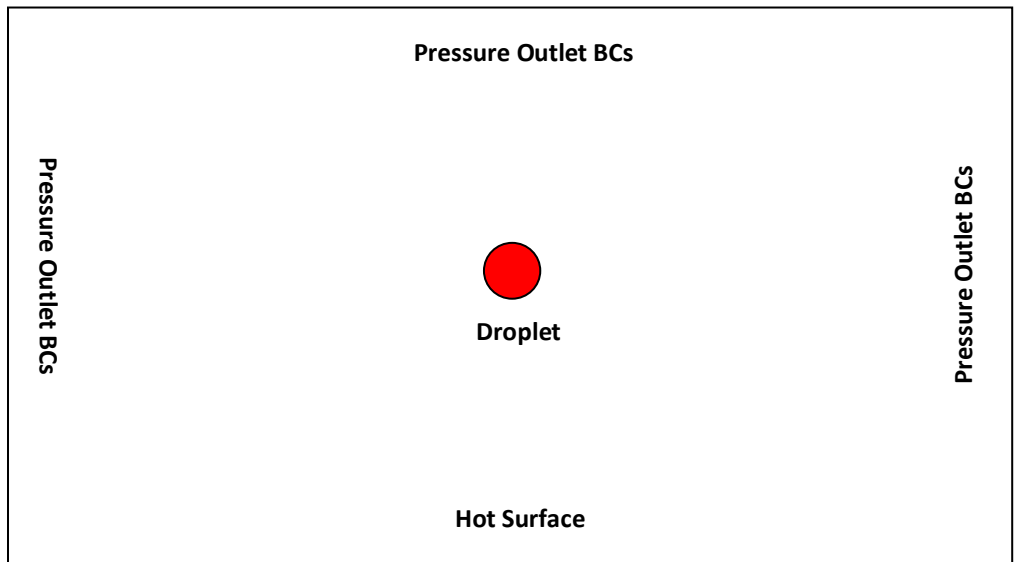

*Figure-3: Sketch of the computational domain and boundary conditions*

The values of material parameters and the operating conditions are given below in Table-1.

*Table-1: Properties of droplet and boundary conditions*

| Surface Used | Copper | Aluminium | Steel |
|---|---|---|---|
| **Roughness (μm)** | 0.718 | 2.582 | 3.527 |
| **Liquid Used** | Methanol, Ethanol, Kerosene, Water | | |
| **Droplet diameter (mm)** | 3.3 | | |
| **Droplet impinging velocity (m/s)** | 2.4, 3.9 & 4.9 | | |
| **Surface temperature (°C)** | 150, 300 & 425 | | |

**k-ω-SST turbulence model:** In 1994 Mentor proposed a 2-equation model designed to yield the best behaviour of k-∈ & k-ω models. It is called non-standard k-ω model. Close to the wall the blending function is 0 (Leading to the standard ω-equation) whereas away from the wall the blending function is 1 which basically corresponds to standard ∈ equation of the k-∈ model.

For the above 2-D geometry rectangular mesh was used and near about 150000 mesh elements had been chosen for the numerical investigations.

3. **RESULTS AND DISCUSSION**

By using the camera we had captured the phenomenon of all three liquid droplets on three different surfaces for the surface three temperatures i.e. 150°, 300° & 425°C and for three impinging velocities i.e. 2.4, 3.9 & 4.9m/s. By changing the height of the injector we had changed their impact velocity. In the following picture i.e. Figure-4, it is shown the result for methanol droplet impinging on three different surfaces and here the time step is 16 ms, surface temperature is 300°C and impinging velocity is 2.4 m/s. After releasing from the needle the droplet took almost 8ms (for that velocity) to reach the surface and after reaching it spread towards its perimeter until it started to break. Spreading ratio is an important phenomenon of droplet after impact it is defined as the ratio of "maximum spreading diameter of droplet after impact" to "the initial droplet diameter". Mathematically it can be written as

$$S = \frac{D_{max}}{D_0} \quad \text{------------------------------------------------- (7)}$$

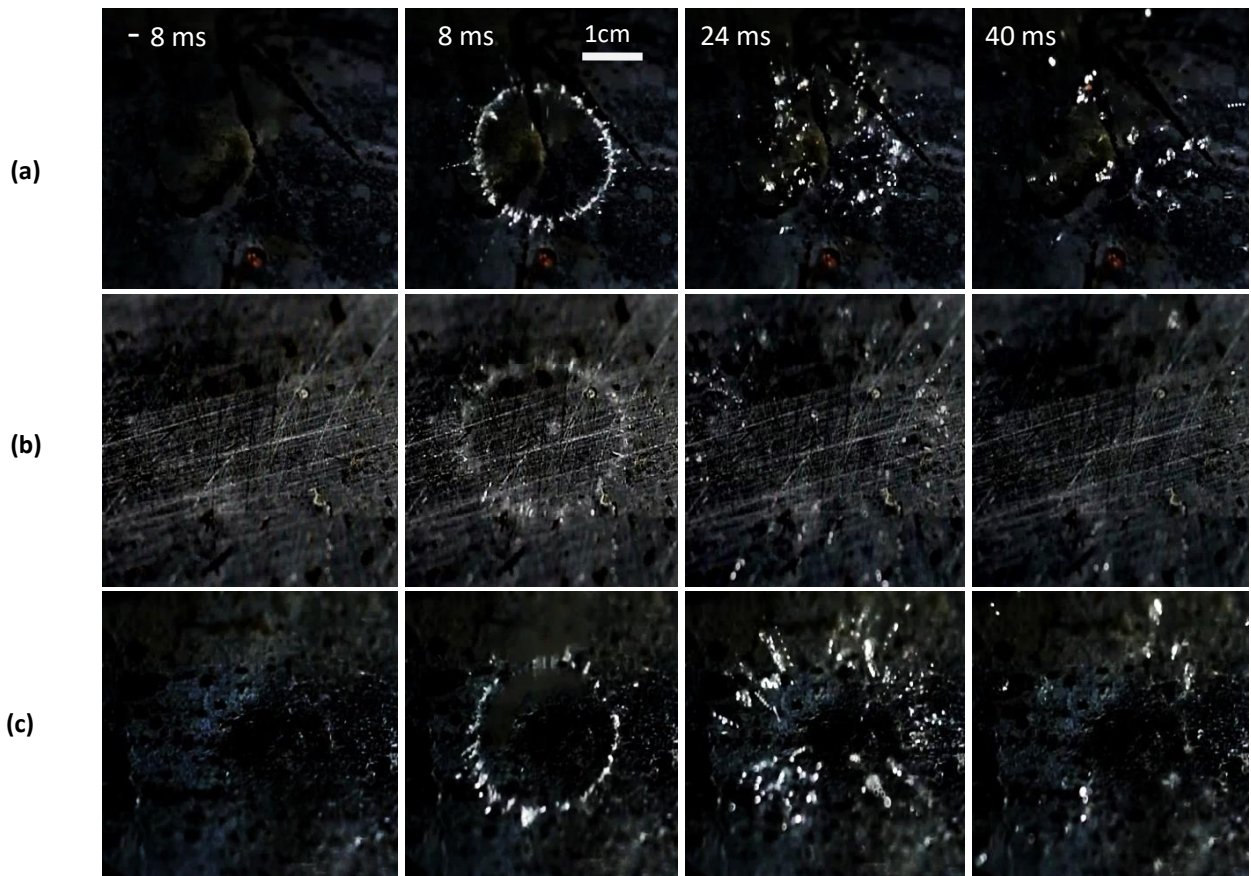

*Figure-4: Methanol droplet on different hot surfaces (2.4m/s 300°C): (a) Copper (b) Aluminium (c) Steel*

Next, we introduced the Weber number to describe the relation between the spreading ration and Weber number. The Weber number for a pure drop, is defined as

$$We = \frac{\rho v^2 D_0}{\sigma} \quad \text{------------------------------------} \quad (8)$$

Secondary droplets were created after droplet spreading on the surface and they depended on initial surface temperature. After impacting the large secondary droplet easily distinguished. For this temperature leidenfrost phenomenon was observed in all three surfaces and also splashing was occurred. Clearly for aluminium surface the secondary droplets escaped faster than the steel and copper. Spreading diameter depends on surface roughness, surface tension & impact energy. The breaking of the droplet looked brighter as a high degree of nucleate boiling was present in the droplet. In the Figure-5 the entire three droplets break up scenario are shown for 425°C. It was observed that the secondary droplets of methanol and ethanol move slower than kerosene. At this high temperature film boiling was occurred.

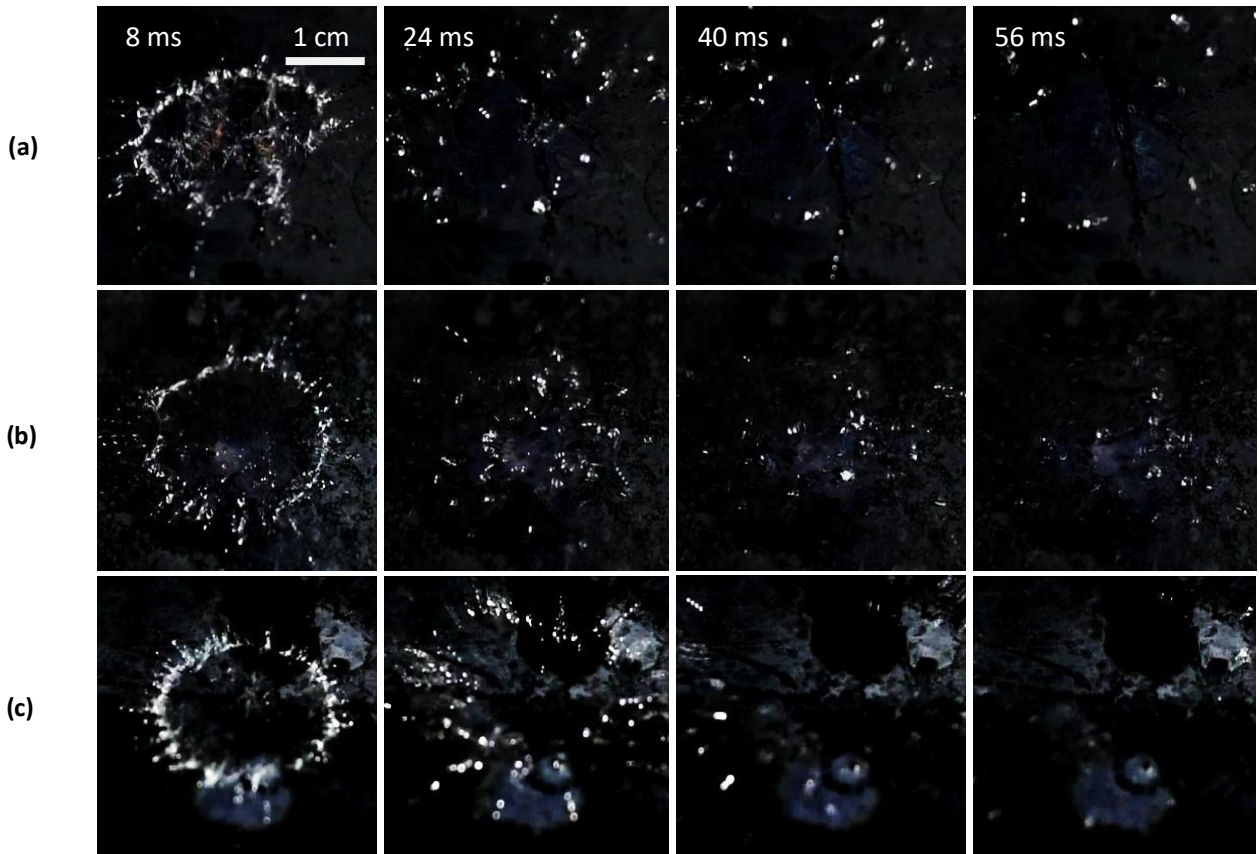

*Figure-5: Different liquid droplets on copper surfaces (2.4m/s 425°C): (a) Methanol (b) Ethanol (c) Kerosene*

The droplets at 4.9m/s impinging speed and 300°C surface temperature is shown in the Figure-6. From the figure we can say the ethanol droplet moves faster and the kerosene droplet moves slower. The fingering boiling was clearly visible for kerosene droplet.

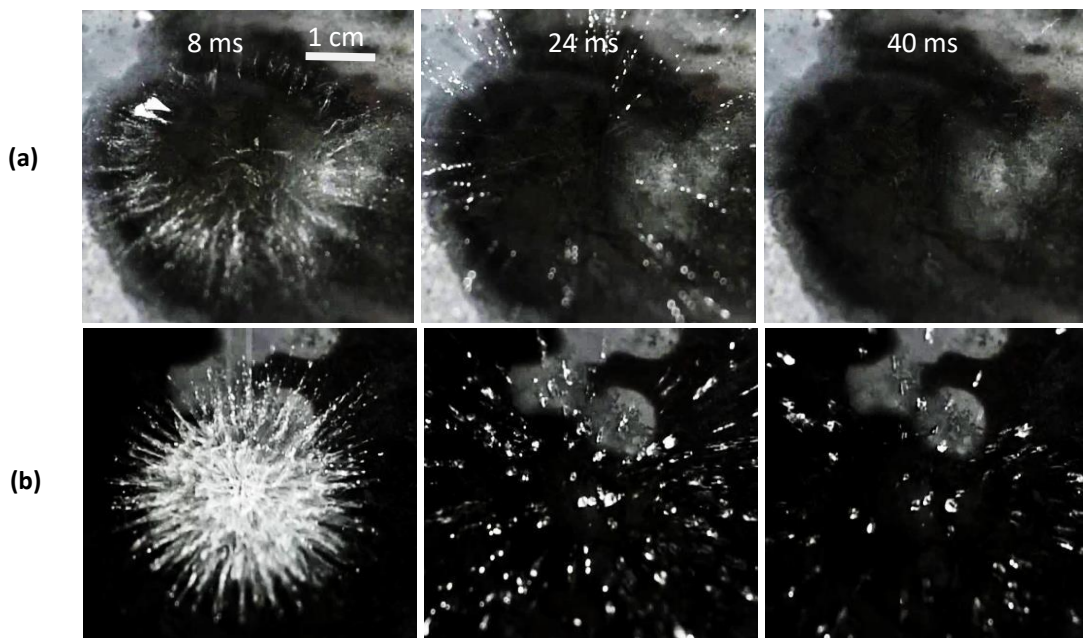

*Figure-6: Different liquid droplets on copper surfaces (4.9m/s 300°C): (a) Ethanol (b) Kerosene*

The experimental values of all the liquids impact with copper surface is given by a table.

*Table-2: Experimental results of all droplets impinging on copper surface at 300°C*

| Droplet (300°C) | Spreading dia (mm) | Density (kg/m$^3$) | Surface tension (N/m) | Drop Diameter (mm) | Velocity (m/s) | S | We |
|---|---|---|---|---|---|---|---|
| Methanol | 21.577 | 792 | 0.023 | 3.3 | 2.4 | 6.538 | 669 |
|  | 26.210 | 792 | 0.023 | 3.3 | 3.9 | 7.942 | 1767 |
|  | 29.231 | 792 | 0.023 | 3.3 | 4.9 | 8.858 | 2789 |
| Ethanol | 21.323 | 789 | 0.022 | 3.3 | 2.4 | 6.462 | 670 |
|  | 27.075 | 789 | 0.022 | 3.3 | 3.9 | 8.205 | 1768 |
|  | 30.205 | 789 | 0.022 | 3.3 | 4.9 | 9.153 | 2791 |
| Kerosene | 20.487 | 786 | 0.026 | 3.3 | 2.4 | 6.208 | 575 |
|  | 24.750 | 786 | 0.026 | 3.3 | 3.9 | 7.500 | 1517 |
|  | 28.175 | 786 | 0.026 | 3.3 | 4.9 | 8.538 | 2395 |

From the above table we had plotted a graph between the spreading ratio and the Weber number for all three surfaces at 300°C. For copper the spreading ratio was varies in between 6.0 - 9.5 for the Weber number range of 500 - 2900. From the Figure-7a we observed that ethanol gave a highest spreading ratio as it had a comparatively lower surface tension value than other two for this the intermolecular force between the fluids particles were very low as a result the liquid could spread better. So ethanol could evaporate faster and the cooling of the surface also went faster where kerosene gave a comparatively slower cooling rate as well as lower evaporation rate as it had a lower spreading ratio.

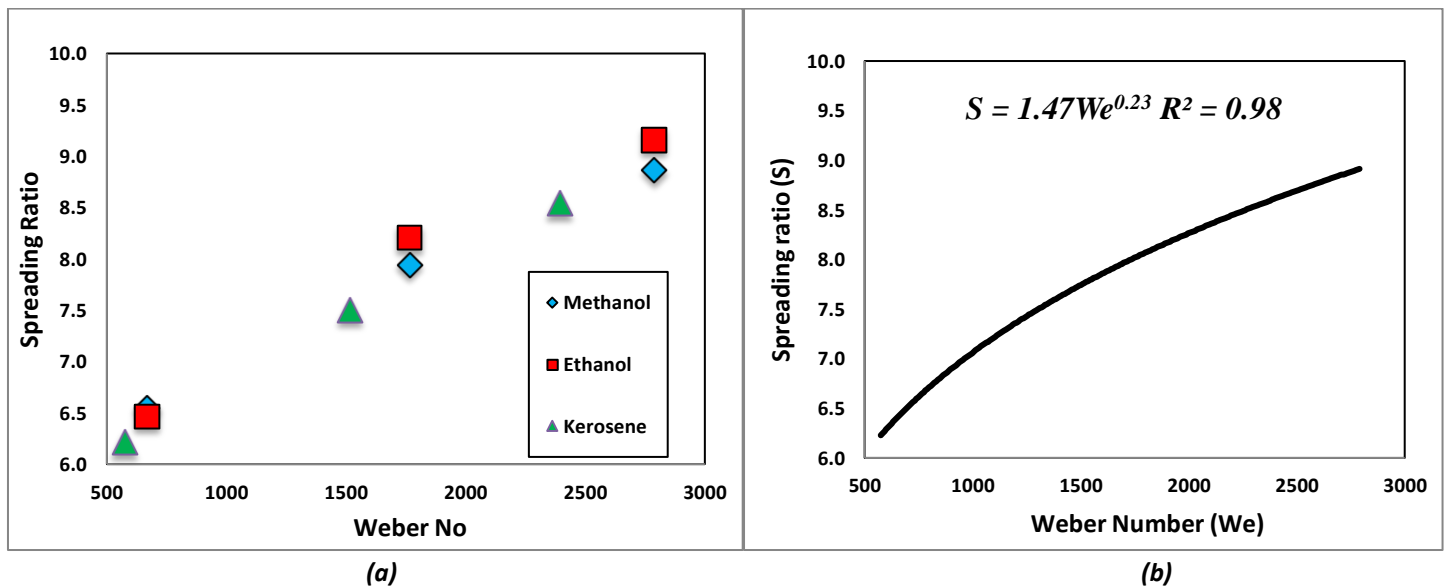

*Figure-7: (a) Spreading ratio vs Weber No, (b) Best fitted curve (Copper Surface)*

The fitted curve equation: $S = 1.47We^{0.23}$, $R^2 = 0.98$

As shown in Figure-7b, by joining the values of the spreading ratio and the weber number we get a best fitted curve which can be written as

$$S = 1.47We^{0.23} \quad \text{------------------------------------- (9)}$$

We can say Spreading ratio is directly proportional to the Weber number. For higher Weber number surface tension value becomes lower and the impact energy becomes higher for this the spreading ratio becomes larger. For Steel and Aluminium surface also it gave almost same result for 300°C surface temperature. In Figure-8 the comparisons graphs in between all there surfaces are plotted. Clearly the copper surface gave better spreading ratio for all the liquid droplets due to less surface roughness.

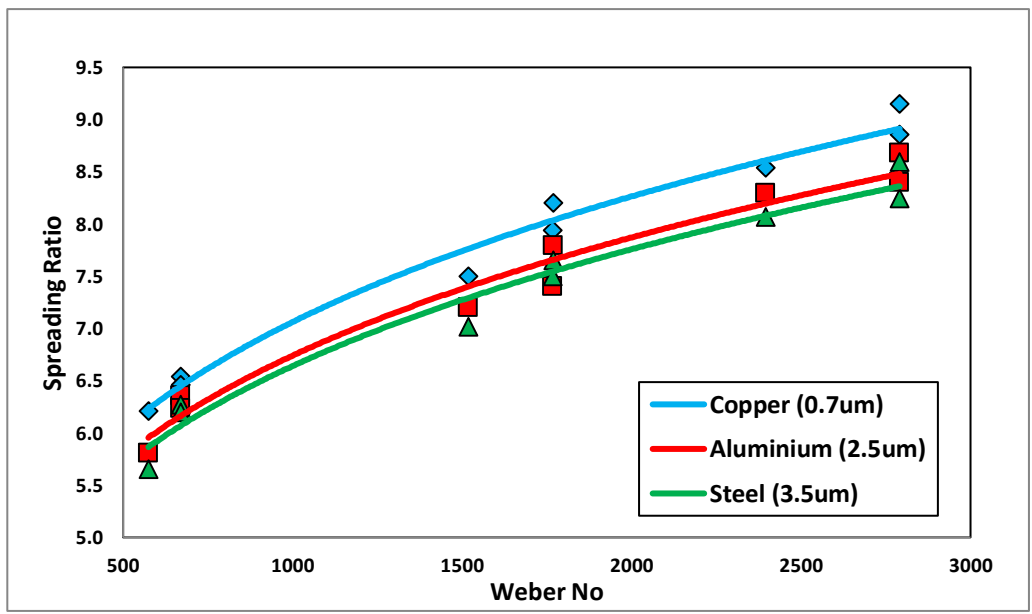

*Figure-8: Spreading ratio vs Weber number at 2.4m/s impinging velocity for different surface*

As the surface roughness value of Steel was higher than rest two it gave lower spreading ratio for the same weber number. From Figure-9a & 9b we can easily say spreading ratio is inversely proportional to surface roughness. This was happened due to the surface friction. For both the impinging velocity kerosene gave lowest spreading ratio. Also for 2.4m/s impinging velocity methanol gave best spreading ratio where for 4.9m/s ethanol gave the best one.

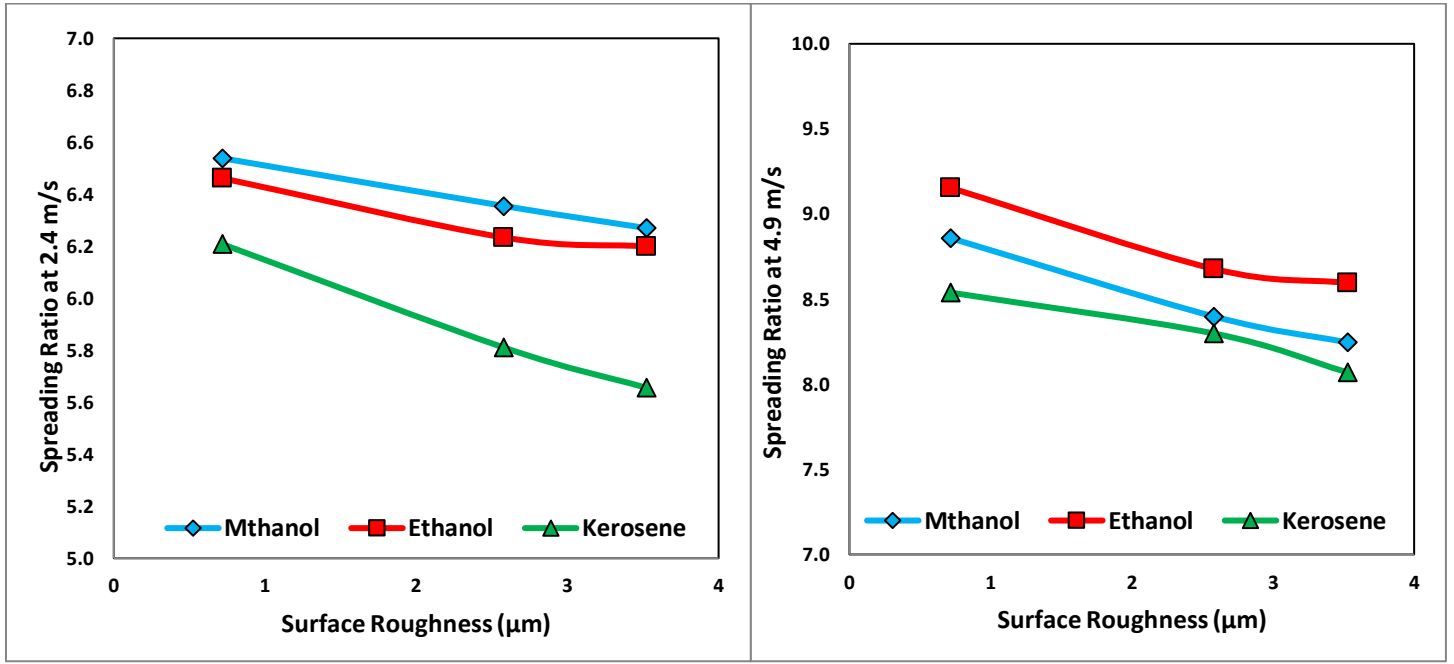

*(a)*           *(b)*

*Figure-9: Surface roughness vs spreading ratio at different impinging velocity (a) 2.4m/s & (b) 4.9m/s*

The interaction stages of methanol droplet are shown in the Figure-10 which was created by computational job. Figure-10a & 10b are the beginning of the interaction and in this stage the impact area was very less but still the heat transfer was effective due to the interaction and the above surface temperature droped rapidly.

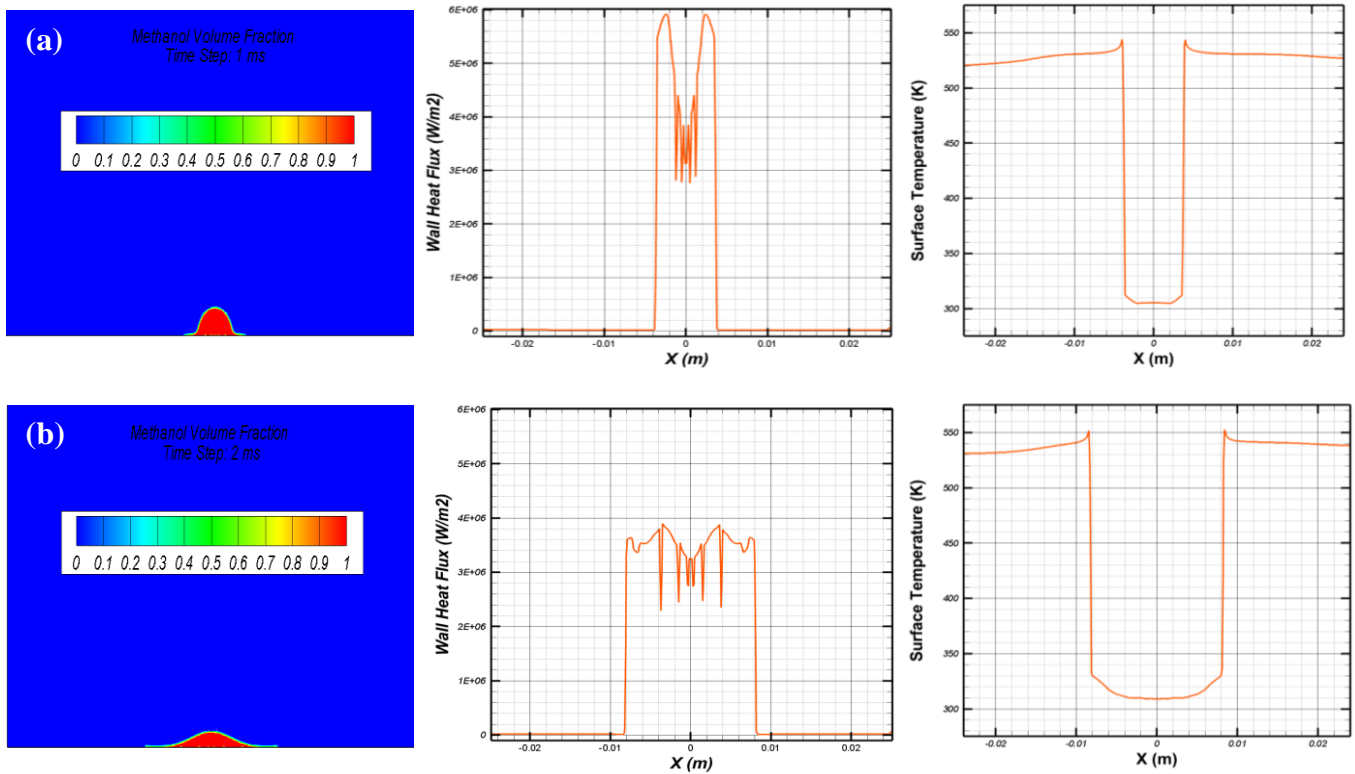

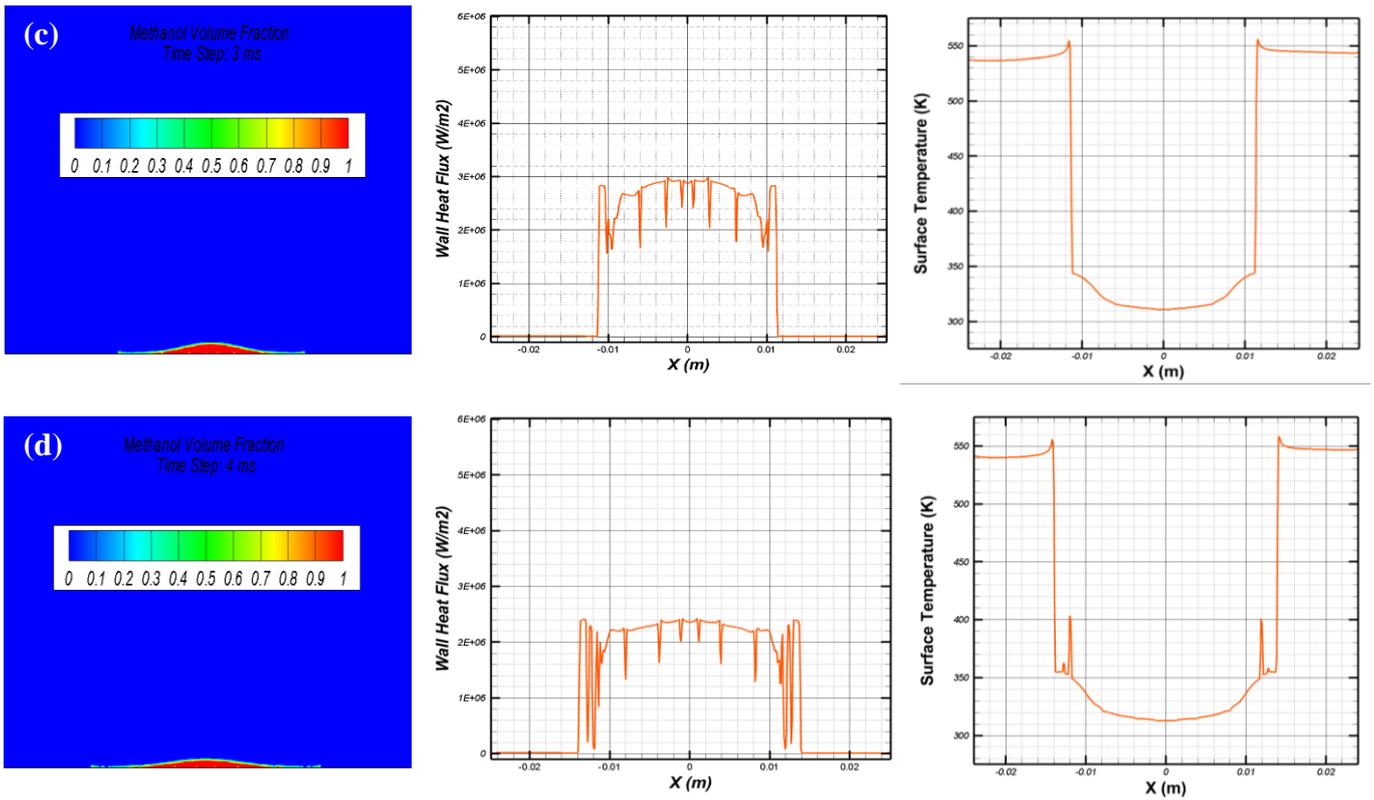

*Figure-10: Hydrodynamic & thermal stages of a single Methanol droplet (v=2.4m/s) interacting with hot Copper plate (300°C)*

In the next stage as shown in Figure-10c & 10d, the covering impact area of the droplet started to expand and for this heat flux decreased and the upper surface temperature increased. Beside this the effectiveness of the cooling was the most efficient at this stage and the droplet started to break. In the Figure-11 we have plotted the graph in a time interval of 50 ms so that we can capture the spreading, breaking and escaping phenomenon of the droplets. We can see from the figure that kerosene droplet takes more time to break and methanol droplets take less time to escape from the initial zone.

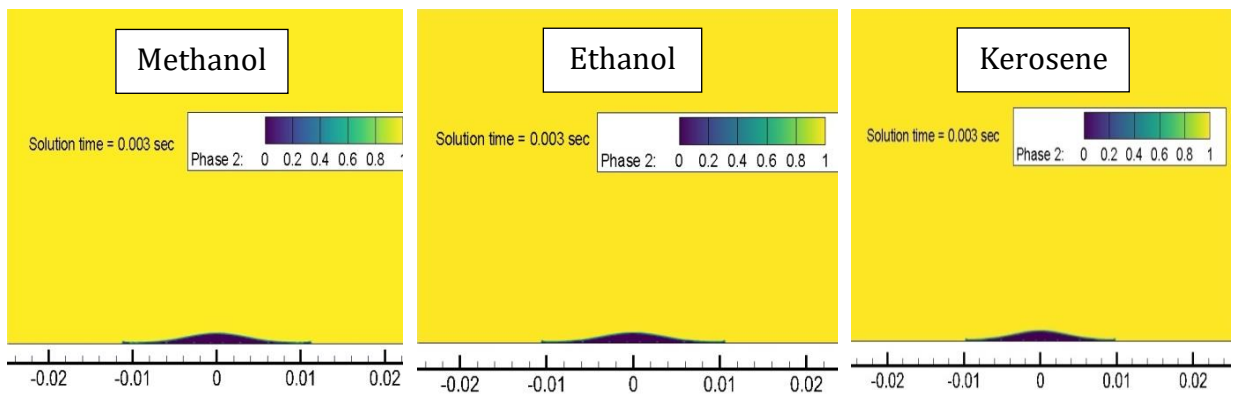

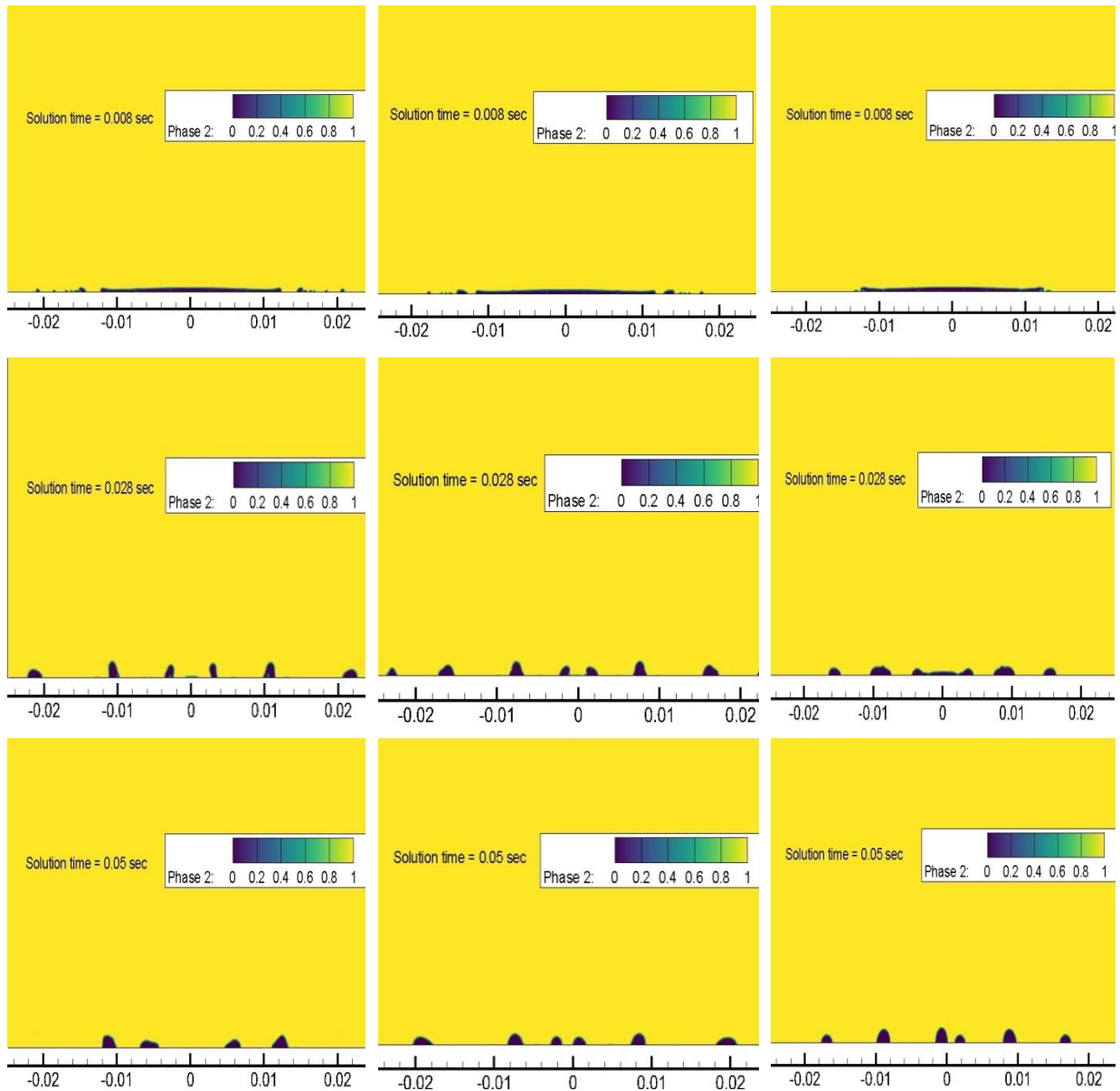

*Figure-11: Methanol, Ethanol, Kerosene droplet on Copper surface at 2.4m/s & 300°C*

In Figure-12, there are three lines for 1ms, 3ms & 5ms and the length denotes the length of the plate. Clearly from the plots we can say for kerosene droplet, the surface temperature reduction is continuously and stable compare to other droplets within a short time interval. The up & down portion of the curve indicates the breaking of the droplet. Compare to other droplets water droplet breaks faster.

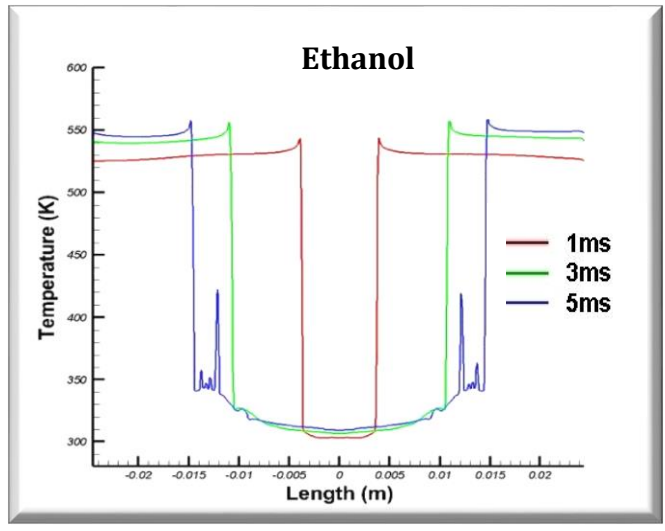
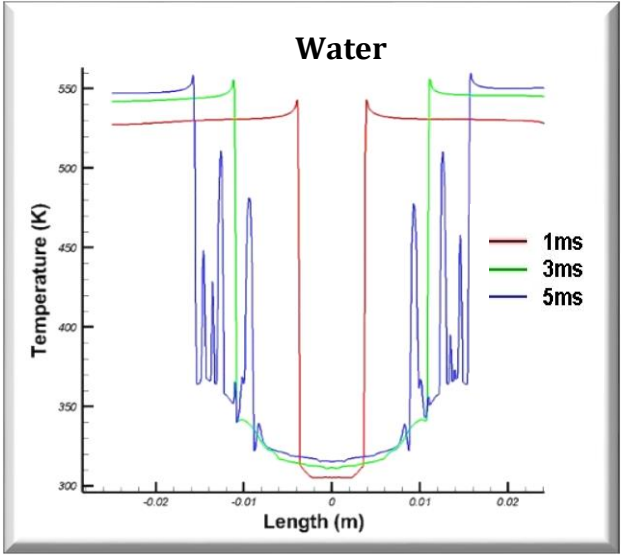
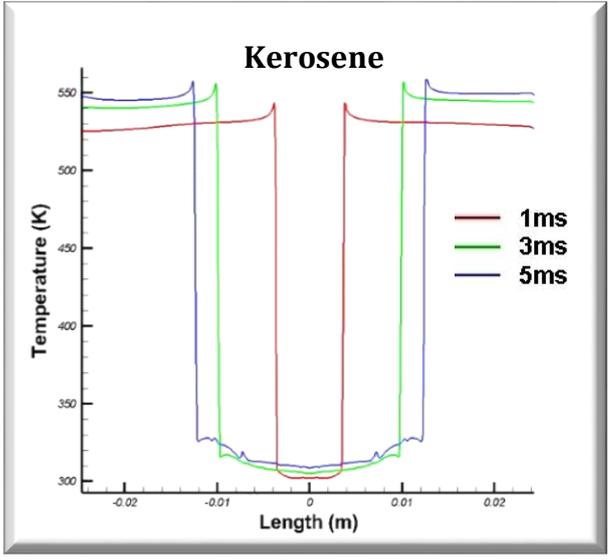

*Figure-12: Copper surface temperature variation for Ethanol, Water & Kerosene (2.4m/s & 300°C)*

From Figure-13 it is observed that kerosene droplet can reduce the upper temperature of the surface in very well manner respect to other three droplets. Basically it is the zoom in view of Figure-12. Also Figure-13 shows us the same scenario and now we can say water droplet can reduce less amount of temperature compare to other.

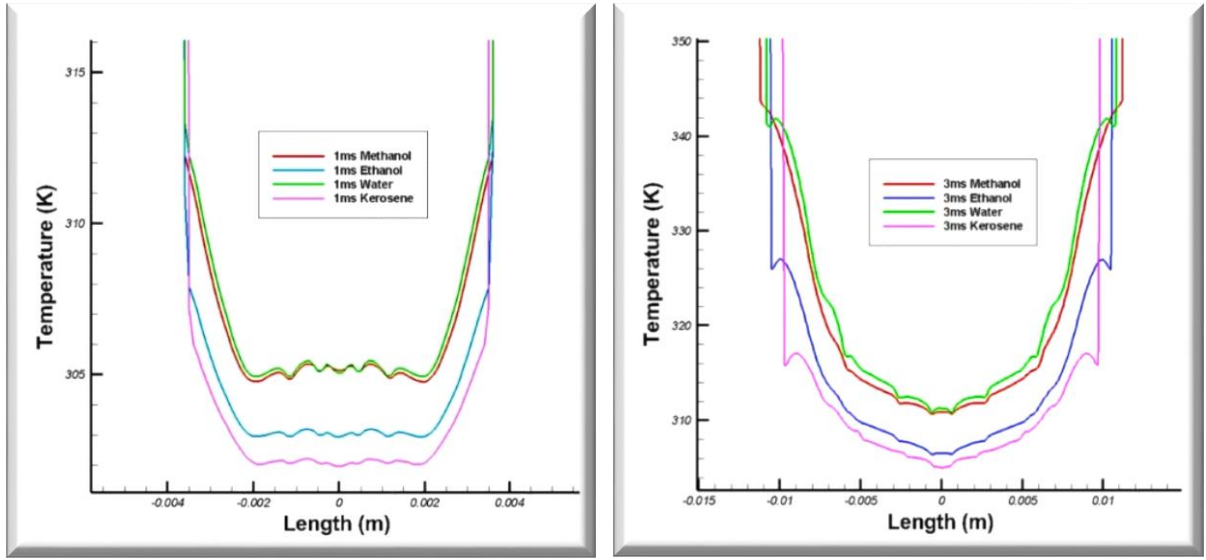

*Figure-13: Methanol, Ethanol, Water & Kerosene droplets on Copper at time step 1ms & 3ms (2.4m/s & 300°C)*

The computational values of all the liquids impact with copper surface is given by a table.

*Table-3: Computational results of all droplets impinging with copper surface at 300°C*

| Droplet (300°C) | Spreading dia (mm) | Density (kg/m$^3$) | Surface tension (N/m) | Drop Dia (mm) | Velocity (m/s) | S | We |
|---|---|---|---|---|---|---|---|
| Methanol | 22.719 | 792 | 0.023 | 3.3 | 2.4 | 6.885 | 669 |
|  | 27.059 | 792 | 0.023 | 3.3 | 3.9 | 8.200 | 1767 |
|  | 32.144 | 792 | 0.023 | 3.3 | 4.9 | 9.741 | 2789 |
| Ethanol | 22.492 | 789 | 0.022 | 3.3 | 2.4 | 6.816 | 670 |
|  | 27.012 | 789 | 0.022 | 3.3 | 3.9 | 8.185 | 1768 |
|  | 32.109 | 789 | 0.022 | 3.3 | 4.9 | 9.730 | 2791 |
| Kerosene | 21.751 | 786 | 0.026 | 3.3 | 2.4 | 6.591 | 575 |
|  | 26.284 | 786 | 0.026 | 3.3 | 3.9 | 7.965 | 1517 |
|  | 30.554 | 786 | 0.026 | 3.3 | 4.9 | 9.259 | 2395 |

We have plotted the experimental and numerical result both and made a comparison in between them. From the Figure-14a we can say methanol gives a higher spreading ratio where kerosene gives a lower spreading ratio at 2.4m/s impinging velocity.

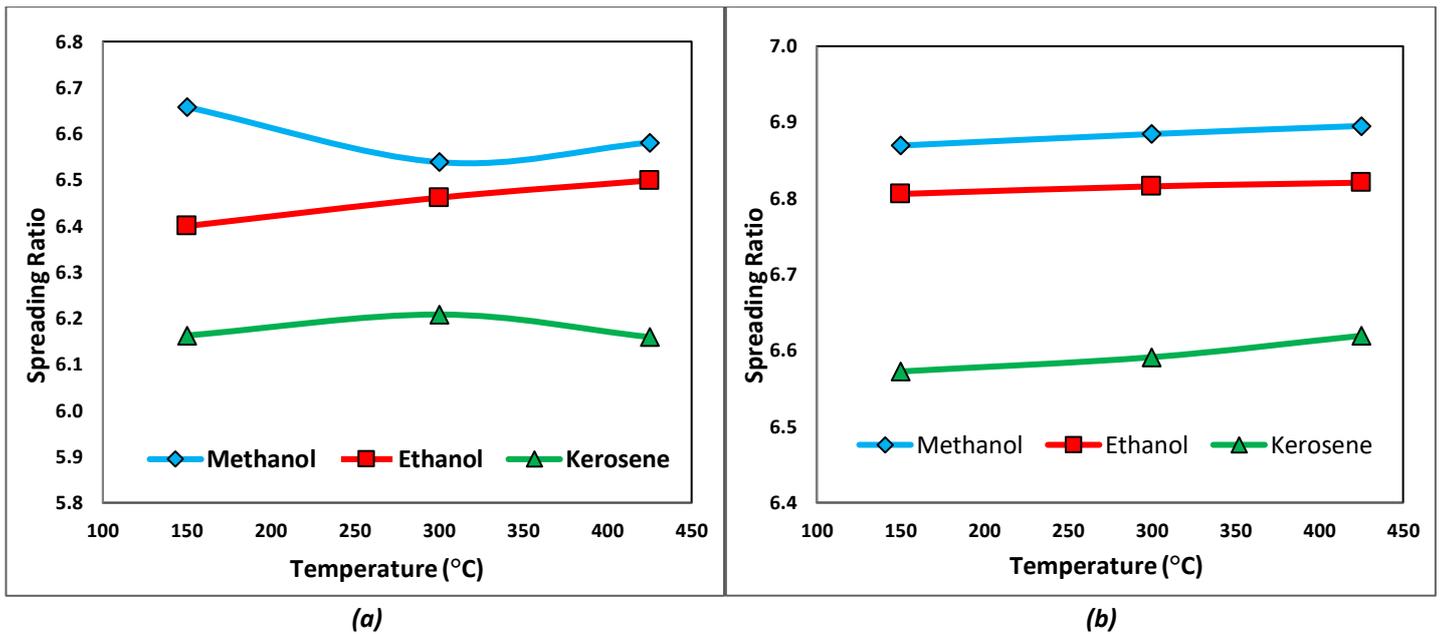

*Figure-14: Spreading ratio vs Surface temperature, (a) Experimental (b) Computational (Copper)*

This is because of the surface tension of the liquid (Surface tension of kerosene is higher than ethanol and methanol). Also in computational case (Figure-14b), kerosene gives a comparatively lower spreading ratio than the other two at 2.4m/s impinging velocity. It is clearly visible in Figure-15 that spreading ratio increases with the increasing of impinging velocity and it looks like linear line. The experimental and the computational results look similar.

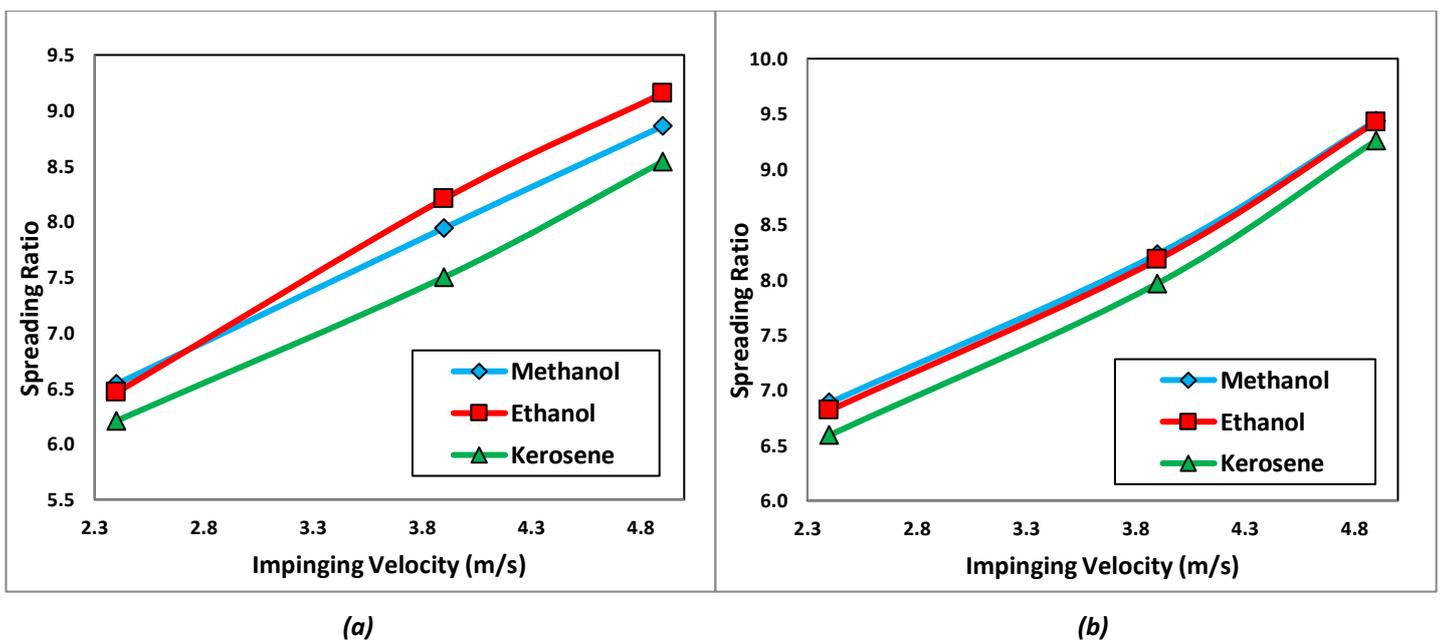

*Figure-15: Spreading ratio vs Impinging velocity, (a) Experimental (b) Computational (Copper)*

The best fitted cure for methanol is shared below in Figure-16.

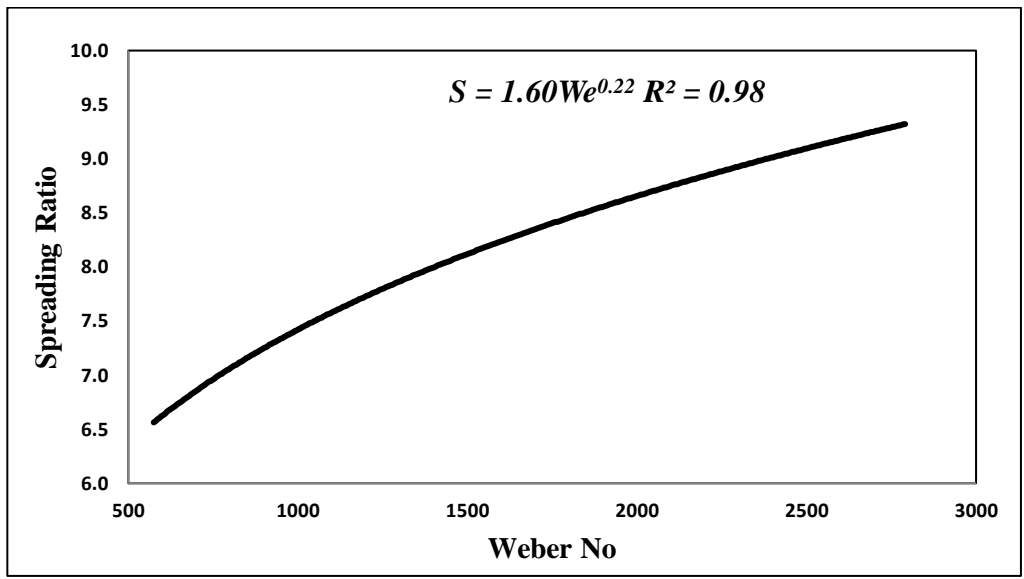

*Figure-16: Spreading ratio vs Weber No; Best fitted curve (Copper Surface)*

The numerical best fitted curve can be written as

$$S = 1.60 We^{0.22} \text{\ -----------------------------------\ (10)}$$

In the Figure-17 two lines are there in which one indicates the computational result and the other indicates the experimental result. Both the curve looks similar but there is an error between them which is lies in between 5-6%. As the roughness values of the surface are not same at all the places for experimental and also for every time the roughness was changing slightly that's why the values of spreading ratio is low compare to computational results.

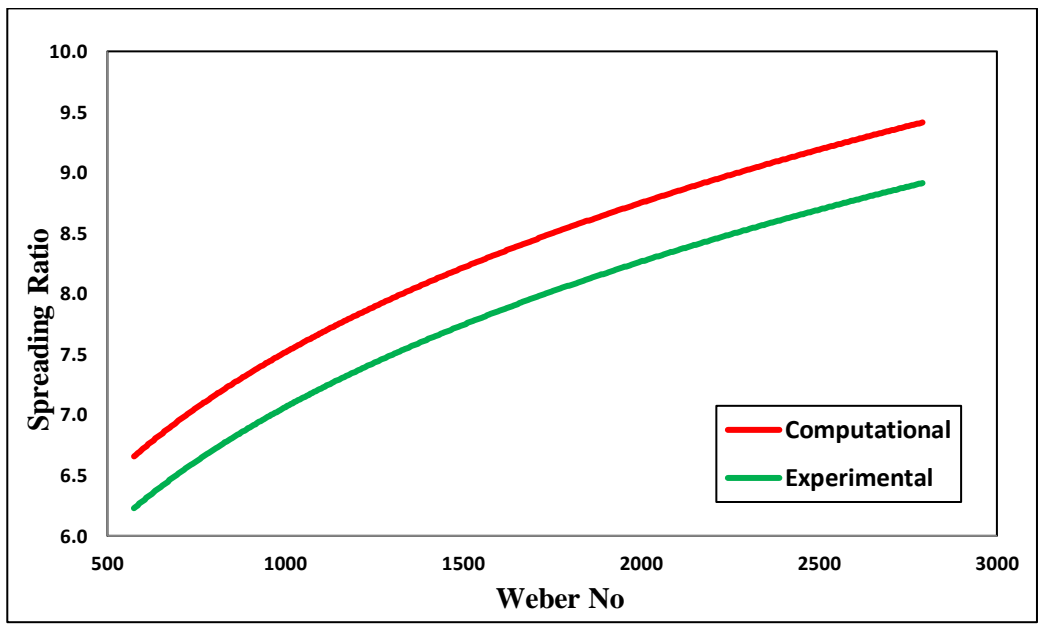

*Figure-17: Experimental results vs Computational results for spreading ratio vs Weber no (Copper surface)*

## 4. CONCLUSIONs

The influence of a single droplet impacting on a hot and solid surface with comparatively high Weber number was analysed in this study. From the result it can be stated that the droplet diameter and the Weber number plays an important role for influencing the heat transfer performance. By morphological observations it is shown that after impacting on the surface secondary droplets were produced and it disintegrated by two mechanisms: inertial and thermally induced breakup. Where the splashing phenomenon was occurred at the time of spreading, there inertial breakup takes place. Where the effects of vapour bubble generation were occurred there thermal breakup observed. The spreading characteristics of the droplet on the surface didn't depend on any parameter rather than the surface roughness. Ethanol gave comparatively larger spreading ratio for any surface. Fingering boiling could be founded in each droplet at 150°C and 300°C. As in this temperature zone nucleate boiling was occurred, and for temperature 425°C film boiling characteristic was observed. When the surface temperature increases prominently past the beginning of nucleate boiling, thermal-induced breakup increases more. Also the surface temperature didn't have any role on influencing the spreading ratio. Weber number is directly proportional to the spreading ratio. Or we can say for a larger value of Weber number spreading ratio also becomes larger. Experimentally we had founded one equation by which we can calculate the spreading ratio value for a particular weber number that is $S = 1.47We^{0.23}$ and for computational the equation becames $S = 1.60We^{0.22}$. Kerosene droplet can reduce the upper temperature of the surface in very well manner respect to other droplets. The deposition rate of heat into the droplet was affected by the thermal diffusivity as it influenced the thermal breakup of droplet. A large explosive boiling force was generated with the increasing of thermal diffusivity and secondary droplet sizes decreased with the higher thermal diffusivity value.

### NOMENCLATURE

| | | |
|---|---|---|
| $K$ | = | Solid surface thermal conductivity |
| $\mu$ | = | Viscosity of the liquid |
| $\dot{\rho}$ | = | Mass source term |
| $\dot{h}$ | = | Energy source term |
| $f_\sigma$ | = | Surface tension source term |
| $g$ | = | Gravitational source term |

$We$ = Weber number
$\rho$ = Density of droplet
$D_0$ = Initial droplet diameter
$\sigma$ = Surface tension of droplet
$V_0$ = Initial droplet velocity
$D_{max}$ = Maximum spreading diameter
$S$ = Spreading ratio

**REFERENCE**

[1] J. Fukai, Z. Zhao, D. Poulikakos, C. M. Megaridis, and O. Miyatake, ''Modeling of the droplet deformation of a liquid droplet impinging upon a flat surface,'' Phys. Fluids A 5, 2588 ~1993!.

[2] J. Fukai, Y. Shiiba, T. Yamamoto, O. Miyatake, D. Poulikakos, C. M. Megaridis, and Z. Zhao, ''Wetting effects on the spreading of a liquid droplet colliding with a flat surface: Experiment and modeling,'' Phys. Fluids 7, 236 ~1995!.

[3] T. Jonas, A. Kubitzek, and F. Obermeier, ''Transient heat transfer and break-up mechanisms of drops impinging on heated walls,'' Experimental Heat Transfer, Fluid Mechanics and Thermodynamics 1887, Proceedings of the Fourth World Conference on Experimental Heat Transfer, Fluid Mechanics and Thermodynamics, Brussels, 1997, p. 1263.

[4] H. Fujimoto and N. Hatta, ''Deformation and rebounding processes of a water droplet impinging on a flat surface above Leidenfrost temperature,'' Trans. ASME, J. Fluids Eng. 118, 142 ~1996!.

[5] D. A. Weiss, ''Periodischer Aufprall monodisperser Tropfen gleicher Geschwindigkeit auf feste Oberfla¨chen,'' Mitteilungen aus dem Max-PlanckInstitut fu¨r Stro¨mungsforschung No. 112, edited by E.-A. Mu¨ller, Go¨ttingen, 1993.

[6] N. Hatta, H. Fujimoto, and H. Takuda, ''Deformation process of a water droplet impinging on a solid surface,'' Trans. ASME, J. Fluids Eng. 117, 394 ~1995!.

[7] A. L. Yarin and D. A. Weiss, ''Impact of drops on solid surfaces: self-similar capillary waves and splashing as a new type of kinematic discontinuity,'' J. Fluid Mech. 283, 141 ~1995!.

[8]  S. Chandra and C. T. Avedisian, ''On the collision of a droplet with a solid surface,'' Proc. R. Soc. London, Ser. A 432, 13 ~1991!.

[9]  A. Karl and A. Frohn , "Experimental investigation of interaction processes between droplets and hot walls" Physics of Fluids, volume 12, number 4, 2000

[10] Cheng W., Zhang W., Chen H., Hu L. "Spray cooling and flash evaporation cooling: The current development and application", Renewable and Sustainable Energy Reviews, 55, 2016, pp. 614–628.

[11] Rini D.P., Chen R.H., Chow L.C.."Bubble behavior and nucleate boiling heat transfer in saturated FC-72 spray cooling", Journal of Heat Transfer, 124, 2002, pp. 63–72.

[12] Liang G., ve Mudawar I." Review of drop impact on heated walls", International Journal of Heat and Mass Transfer 106, 2017, pp. 103–126

[13] Liang G., ve Mudawar I." Review of spray cooling – Part 1: Single-phase and nucleate boiling regimes, and critical heat flux", International Journal of Heat and Mass Transfer 115, 2017, pp. 1174–1205.

[14] Liang G., ve Mudawar I." Review of spray cooling – Part 2: High temperature boiling regimes and quenching applications", International Journal of Heat and Mass Transfer 115, 2017, pp. 1206–1222.

[15] Yarin, A.L. Drop Impact Dynamics: Splashing, Spreading, Receding, Bouncing . . . . Annu. Rev. Fluid Mech. 2006, 38, 159–192.

[16] Khojasteh, D.; Kazerooni, M.; Salarian, S.; Kamali, R. Droplet impact on superhydrophobic surfaces: A review of recent developments. J. Ind. Eng. Chem. 2016, 42, 1–14.

[17] Sara Moghtadernejad, Christian Lee and Mehdi Jadidi. (2020), "An Introduction of Droplet Impact Dynamics to Engineering Students", Fluids Journal.

[18] Worthington, A.M. On the forms assumed by drops of liquids falling vertically on a horizontal plate. Proc. R. Soc. Lond. 1876, 25, 261–272.

[19] Yarin, A.L. Drop impact dynamics: Splashing, spreading, receding, bouncing…Annu. Rev. Fluid Mech. 2006, 38, 159–192.

[20] Alizadeh, A.; Bahadur, V.; Zhong, S. Temperature dependent droplet impact dynamics on flat and textured surfaces. App. Phy.Lett. 2012, 11, 7699.

[21] Jin, Z.; Zhang, H.; Yang, Z. Experimental investigation of the impact and freezing processes of a water droplet on an ice surface. Int. J. Heat Mass Transf. 2017, 109, 716–724.

[22] Jin, Z.; Sui, D.; Yang, Z. The impact, freezing, and melting processes of a water droplet on an inclined cold surface. Int. J. Heat Mass Transf. 2015, 90, 439–453.

[23] Sehgal, B.R., Nourgaliev, R.R., Dinh, T.N. "Numerical simulation of droplet deformation and break-up by LatticeBoltzmann method". Prog. Nucl. Energ. 34 (4), 1999, pp. 471– 488

[24] Fakhari, A. ve Rahimian, M.H. "Investigation of deformation and breakup of a falling droplet using a multiple-relaxation time Lattice Boltzmann method". J. Comput. Fluids 40, 2011, pp. 156–171.

[25] Xiong W., Cheng P. "3D lattice Boltzmann simulation for a saturated liquid droplet at low Ohnesorge numbers impact and breakup on a solid surface surrounded by a saturated vapor". Computers and Fluids168, 2018, pp. 130–143.

[26] Liu C., Shen M., Wu J."Investigation of a single droplet impact onto a liquid film with given horizontal velocity". European Journal of Mechanics / B Fluids 67, 2018, pp. 269– 279.

[27] Beni M.S., Zhao J., Yu K.N. "Investigation of droplet behaviors for spray cooling using level set Method". Annals of Nuclear Energy 113, 2018, pp. 162–170.

[28] M. Bussmann, S. Chandra, J. Mostaghimi, Phys. Fluids 12: 3121-3132 (2000).

[29] L.R. Villegas, S. Tanguy, G. Castanet, O. Caballina, F. Lemoine, Int. J. Heat Mass Tran. 104: 1090-1109 (2017).

[30] Bernardin J.D., Stebbins C.J, Mudawar I. "Mapping of impact and heat transfer regimes of water drops impinging on a polished surface", Int. J. Heat Mass Transfer 40, 1997, pp. 247–267.

[31] Erkan N., Kawakami T., Madokoro H., Chai P., Ishiwatari Y., Okamoto K. "Numerical simulation of droplet deposition onto a liquid film by VOF–MPS hybrid method", Journal of Visualization, 18, 2015, pp. 381–391.